\documentclass[twocolumn,nofootinbib,preprintnumbers,amsmath,amssymb,superscriptaddress]{revtex4}
\usepackage{epsfig}
\pdfoutput=1

\usepackage[mathletters]{ucs}
\usepackage[utf8x]{inputenc}

\usepackage{amsmath,amssymb,amsthm,amsfonts,mathtools}
\usepackage{mathrsfs}
\usepackage{slashed}
\usepackage{graphicx}
\usepackage{subfigure}
\usepackage[dvipsnames]{xcolor}
\usepackage{rotating}
\usepackage{dsfont}
\usepackage{setspace}
\usepackage{verbatim}
\usepackage{fancyhdr}
\usepackage
[pdftitle={Towards Novel Insights in Lattice Field Theory with Explainable Machine Learning},
pdfsubject={Towards Novel Insights in Lattice Field Theory with Explainable Machine Learning}]
		{hyperref}
\usepackage{cleveref}
\usepackage{tabularx}

\newcommand{\avg}[1]{\langle #1 \rangle}

\begin{document}
\title{Towards Novel Insights in Lattice Field Theory\\
	   with Explainable Machine Learning} 
\author{Stefan Bl\"ucher}
\affiliation{Institut f\"ur Theoretische
  Physik, Universit\"at Heidelberg, Philosophenweg 16, 69120
  Heidelberg, Germany}

\author{Lukas Kades}
\affiliation{Institut f\"ur Theoretische
	Physik, Universit\"at Heidelberg, Philosophenweg 16, 69120
	Heidelberg, Germany}

\author{Jan M.~Pawlowski}
\affiliation{Institut f\"ur Theoretische
  Physik, Universit\"at Heidelberg, Philosophenweg 16, 69120
  Heidelberg, Germany}
\affiliation{ExtreMe Matter Institute EMMI, GSI,
  Planckstra{\ss}e 1, D-64291 Darmstadt, Germany}

\author{Nils Strodthoff}
\affiliation{Fraunhofer Heinrich Hertz Institute, 10587 Berlin, Germany}

\author{Julian M.~Urban}
\affiliation{Institut f\"ur Theoretische
	Physik, Universit\"at Heidelberg, Philosophenweg 16, 69120
	Heidelberg, Germany}

\begin{abstract}
Machine learning has the potential to aid our understanding of phase
structures in lattice quantum field theories through the statistical
analysis of Monte Carlo samples. Available algorithms, in particular
those based on deep learning, often demonstrate remarkable performance
in the search for previously unidentified features, but tend to lack
transparency if applied naively. To address these shortcomings, we
propose representation learning in combination with interpretability
methods as a framework for the identification of observables. More
specifically, we investigate action parameter regression as a pretext
task while using layer-wise relevance propagation (LRP) to identify
the most important observables depending on the location in the phase
diagram. The approach is put to work in the context of a scalar Yukawa
model in (2+1)d. First, we investigate a multilayer perceptron to
determine an importance hierarchy of several predefined, standard
observables. The method is then applied directly to the raw field
configurations using a convolutional network, demonstrating the
ability to reconstruct all order parameters from the learned filter
weights. Based on our results, we argue that due to its broad
applicability, attribution methods such as LRP could prove a useful
and versatile tool in our search for new physical insights. In the
case of the Yukawa model, it facilitates the construction of an
observable that characterises the symmetric phase.
\end{abstract}

\maketitle

\section{Introduction}

Lattice simulations of quantum field theories have proven essential
for the theoretical understanding of fundamental interactions from
first principles, perhaps most prominently so in quantum
chromodynamics. However, an in-depth understanding of the emergent
dynamics is often difficult. In cases where such an understanding
remains elusive, it may be instructive to search for so far
unidentified structures in the data to better characterise the
dynamics.

In this quest towards new physical insight, we turn to machine
learning (ML) approaches, in particular from the subfield of deep
learning \cite{LeCun2015}. These methods have proven capable of
efficiently identifying high-level features in a broad range of data
types---in many cases, such as speech or image recognition, with
spectacular success
\cite{graves2013speech,NIPS2012_4824,NIPS2015_5638,he2016deep}.
Accordingly, there is growing interest in the lattice community to
harness the capabilities of these algorithms, both for high energy
physics and condensed matter systems. Applications include predictive
objectives, such as detecting phase transitions from lattice
configurations, as well as generative modeling \cite{Wang2016,
	Broecker2016, Pang:2016vdc, Wetzel2017a, Wetzel2017,
	Cristoforetti:2017naf, Chng2017, Hu2017, Liu2017, Carleo2017,
	Nieuwenburg2017, Ponte2017, Carrasquilla2017, Morningstar2017,
	Tanaka2017, 10.7566/JPSJ.86.063001, Huang2017, Liu2017a, Wu_2019,
	Zhou_2019, Shanahan2018, Suchsland2018, Urban2018a, Noe2019,
	Nicoli2019comment, Albergo2019, Kades2019, Kashiwa2019, Liu2019,
	Casert2019, Rzadkowski2019, Nicoli2019asymptotically,
	10.1093/ptep/ptz082, Greplova:2019fxu}. We recommend \cite{Mehta2019}
as an introduction to ML for physicists and \cite{Carleo2019} as a
general review for ML applications in physics.

One ansatz for the identification of relevant observables from lattice
data is through representation learning, i.e.\ by
training on a pretext task. The rationale behind this approach is that
the ML algorithm learns to recognise patterns which can be leveraged to
construct observables from low-level features that characterise
different phases. However, solving a given task does by itself not
lead to physical insights, since the inner structure of the algorithm
typically remains opaque. This issue can at least partially be
resolved by the use of ``explainable AI'' techniques, which have
recently attracted considerable interest in the ML community
and beyond. In this work, we focus on layer-wise relevance propagation
(LRP) \cite{BachPLOS15}. It is one of several popular post-hoc
attribution methods that propagate the prediction back to the input
domain, thereby highlighting features that influence the algorithm
towards/against a particular classification decision.

We test this approach in the context of Yukawa theory in (2+1)
dimensions, using inference of an action parameter as a pretext task
in order to identify relevant observables. In a first step, we
demonstrate that this is at least partially possible by training a
multilayer perceptron (MLP) on a set of standard observables. Here, we
show that the relevance of features in different phases, as determined
by LRP, agrees with physical expectations. We benchmark our results
with a similar method based on random forests. Subsequently, we
demonstrate that the action parameter can be inferred directly from
field configurations using a convolutional neural network (CNN). We
use LRP to identify relevant filters and discuss how these align with
physical knowledge. This also allows us to construct an observable
that appears to be a distinctive feature of the paramagnetic phase.

The paper is organised as follows. In \Cref{sec:yukawa} we briefly
review scalar Yukawa theory on the lattice and define important
quantities. \Cref{sec:lrp} serves as an introduction to the topic of
explainable AI and discusses LRP in order to convey the rationale
behind our approach. Numerical results for the MLP and a random forest
benchmark are presented in \Cref{sec:exp}. In \Cref{sec:extract} we
conduct an analysis of the CNN and subsequently demonstrate how all
order parameters, as well as the aforementioned observable relevant
for the paramagnetic phase, can be extracted from the filters. We
discuss our findings and possible future work in
\Cref{sec:conclusions}.

\section{Yukawa theory}
\label{sec:yukawa}

We consider a scalar Yukawa model defined on a (2+1)d cubic lattice
with periodic boundary conditions. The theory is comprised of a
real-valued scalar field with quartic self-interaction coupled
to Dirac fermions. The action for the scalar field can be cast into
the following dimensionless form,

\begin{align}\nonumber
S_{\text{KG}}[\phi] &=
\sum_{n\in\Lambda}\Biggr[-2\kappa\sum_{\mu=1}^{d}\phi(n)\phi(n+\hat{\mu}) \\
&\hspace{2cm}+ (1- 2\lambda)\phi(n)^2 + \lambda\phi(n)^4\Biggr]\,,
\label{eq:LatticeKG}
\end{align}
where $\Lambda$ denotes the set of all lattice sites. Here, $\kappa$
is called the hopping parameter and $\lambda$ takes the role of the
coupling constant.

In order to ensure positivity of the partition function, one needs a
minimum of two degenerate fermion flavors. Due to their bilinear
contributions to the action, the fermionic d.o.f.\ can be integrated
out, yielding the determinant of the discretised Dirac operator,

\begin{align}\nonumber
D_{nm}[\phi] &= \sum_{\mu = 1}^{d}\eta_\mu(n) \frac{\delta(n - m +
	\hat{\mu}) - \delta(n - m - \hat{\mu})}{2} \\
& \hspace{1.8cm}+ \,\delta(n - m)\big(M_f + g\,\phi(n)\big)\,,
\label{eq:LatticeDiracMatrix}
\end{align}
as a multiplicative contribution to the statistical weight. The
Euclidean Dirac $\gamma$-matrices are absorbed by the staggered
transformation, yielding the scalars $\eta_\mu(n)$ that mix the
spatial and spinor degrees of freedom. They are given by $\eta_1(n)=1$
and $\eta_l(n) = (-1)^{n_1}\cdots(-1)^{n_{l-1}}$. $M_f$ denotes the
fermion mass and $g$ is the Yukawa coupling to the bosonic field. The
expectation value of an observable $\mathcal{O}$ can then be expressed
as the path integral

\begin{align}
\langle \mathcal{O} \rangle = \frac{1}{\mathcal{Z}} \int \mathcal{D}\phi \, \det(D^T D) \, \exp(-S_{KG}) \ \mathcal{O}(\phi) \,,
\end{align}
where $\mathcal{Z}$ denotes the partition function.
\begin{figure}
	\includegraphics[width=.97\linewidth]{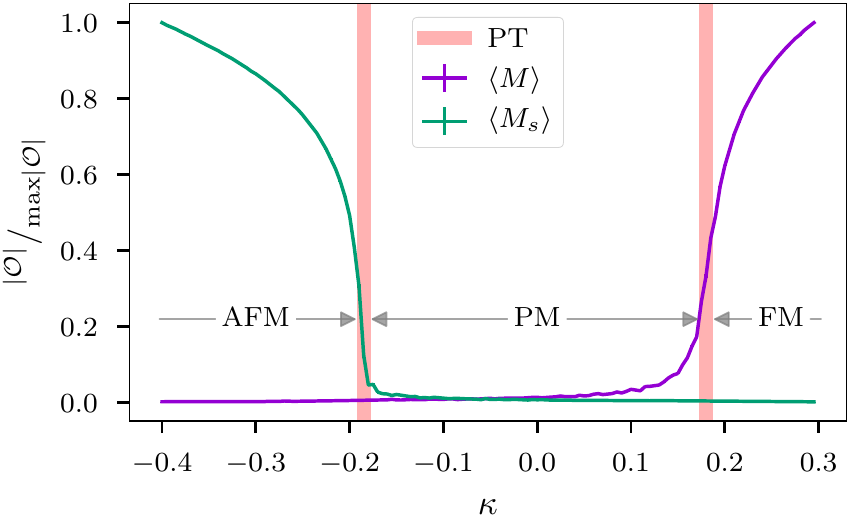}
	\caption{Slice of the phase diagram for fixed Yukawa coupling
	$g=0.25$ using normalised values of $\avg{M}$ and $\avg{M_s}$. Phase
	transitions are highlighted by the shaded bars. We distinguish an
	antiferromagnetic (AFM), a paramagnetic (PM) and a ferromagnetic (FM)
	phase.}
	\label{fig:phase}
\end{figure}
Important observables characterising phases and critical
phenomena in scalar $\phi^4$-theory include the magnetisation, 
\begin{align}
M = \frac{1}{|\Lambda|}\sum_{n \in \Lambda} \phi(n)\,,
\label{eq:Magnetisation}
\end{align}
as well as the staggered magnetisation

\begin{align}
M_s = \frac{1}{|\Lambda|}\sum_{n \in \Lambda} (-1)^{n_1 + \cdots + n_d}
\phi(n) ,
\label{eq:StaggeredMagnetisation}
\end{align}
which is relevant for negative $\kappa$. The scalar part of
\Cref{eq:LatticeKG} features the additional staggered symmetry
\begin{align}
\kappa \mapsto -\kappa	\qquad \text{and} \qquad \phi(n) \mapsto (-1)^n\phi(n),
\label{eq:StaggeredSymmetry}
\end{align}
which connects both magnetisations. The fermionic part explicitly
breaks this symmetry.

A slice of the phase diagram at fixed Yukawa coupling is shown in
\Cref{fig:phase}. The theory exhibits an interesting structure, with
two broken phases of ferromagnetic (FM) and antiferromagnetic (AFM)
nature, where $M$ and $M_s$ respectively acquire non-zero expectation
values. They are separated by a symmetric, paramagnetic (PM) phase,
where both quantities vanish.

We also consider the connected two-point correlation function

\begin{align}
G_c(n, m) = \avg{\phi(n)\phi(m)} - \avg{\phi(n)}\avg{\phi(m)}\,.
\label{eq:ConnectedCorrelation}
\end{align}
While the expectation value of the magnetisation can be estimated from
a single field configuration at reasonable lattice sizes, signals of
n-point correlators are naturally much more suppressed due to
statistical noise and cannot be reasonably approximated from one
sample. Therefore, we introduce the time-sliced correlator $G_c(t)$,
which is defined by

\begin{equation}
G_{c}(t)=\frac{1}{|\Lambda|} \sum_{\vec{n}} G_{c}((t, \vec{n}), (0,\vec 0))\,,
\end{equation}
where the sum runs over spacelike components. It measures correlations
only in the temporal direction, which leads to a better
signal-to-noise ratio due to the averaging procedure.

Some aspects of the derivation and simulation are given in
\Cref{app:simulation}. For a comprehensive treatment of Yukawa theory
on the lattice, we recommend \cite{Montvay1994}.

\section{Insights from Explainable AI}
\label{sec:lrp}

Simple methods from statistics and ML often lack the capability to
model complex data, whereas sophisticated algorithms typically tend to
be less transparent. A commonly used example is principal component
analysis (PCA). It has been successfully applied to the extraction of
(albeit already known) order parameters for various systems
\cite{Wang2016, Wetzel2017a, Hu2017}. However, its linear structure
prohibits the identification of complex non-linear features, e.g.\
Wilson loops in gauge theories. Hence, we require tools capable of
modeling non-linearities, such as deep neural networks
\cite{Carrasquilla2017}. They allow for a more comprehensive treatment
of complex systems, which has been demonstrated e.g.\ for fermionic
theories in \cite{Broecker2016, Chng2017}. The approach also enables
novel procedures, such as learning by confusion and similar
techniques, to locate phase transitions in a semi-supervised manner
\cite{Nieuwenburg2017, Rzadkowski2019}. For lattice QCD, action
parameters can be extracted from field configurations
\cite{Shanahan2018}. Overall, deep learning tools seem particularly
well-suited to grasp relevant information about quantum field dynamics
in a completely data-driven approach, by learning abstract internal
representations of relevant features.

However, their lack of transparency is frequently a major drawback of
using such methods, which prohibits access to and comprehension of
these representations. A unified understanding of how and what these
architectures learn, and why it seems to work so well in a wide range
of applications, is still pending. To better understand the processes
behind neural network-driven phase detection in lattice models,
multiple proposals have been made, such as pruning \cite{Wetzel2017,
	Kashiwa2019, Suchsland2018}, utilising (kernel) support vector
machines \cite{Ponte2017, Liu2019}, and saliency maps
\cite{Casert2019}. Interpretability is also investigated for other
applications in theoretical physics, e.g.\ by employing twin neural
networks \cite{wetzel2020discovering}.

Also, in the broader scope of ML research, there has been growing
interest in interpretability approaches, most of them focusing on
post-hoc explanations for trained models. So-called attribution
methods typically assign a relevance score to each of the input
features that quantifies which features the classifier was
particularly sensitive to, or influenced the algorithm towards/against
an individual classification decision. In the domain of image
recognition, such attribution maps are typically visualised as
heatmaps overlaying the input image. The development of attribution
algorithms is a very active field of research in the ML community.
Therefore, we refer to dedicated research articles for more in-depth
treatments \cite{DBLP:journals/corr/abs-1711-06104,MonDSP18}. Very
broadly, the most important types of such local interpretability
methods can be categorised as: 1.~Gradient-based, such as saliency
maps \cite{Simonyan2013} obtained by computing the derivative of a
particular class score with respect to the input features or
integrated gradients \cite{Sundararajan2017AxiomaticAF}.
2.~Decomposition-based, such as layer-wise relevance propagation (LRP)
\cite{BachPLOS15} or DeepLift \cite{pmlr-v70-shrikumar17a}. 3.
Perturbation-based, as in \cite{DBLP:conf/eccv/ZeilerF14},
investigating the change in class scores when occluding parts of the
input features.

\begin{figure}
	\includegraphics[width=.97\linewidth]{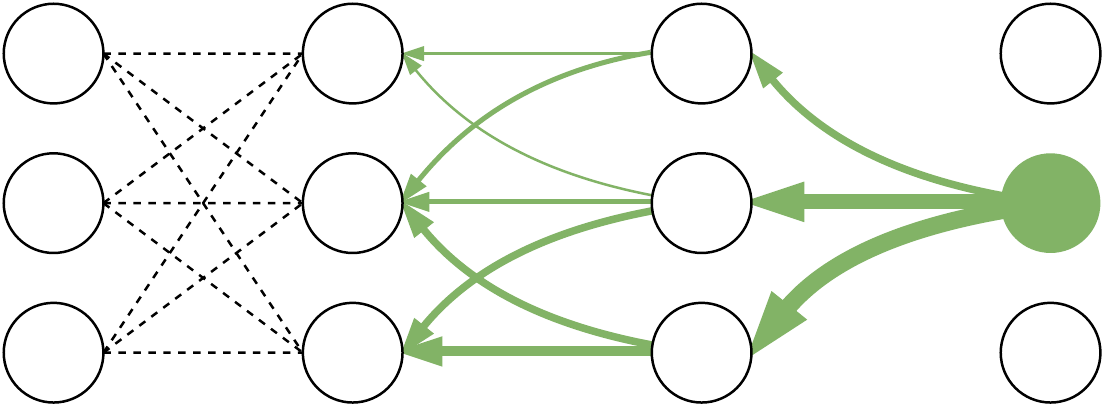}
	\caption{Sketch of LRP through the last two layers of a
		classification network that predicts one-hot vectors. Relevance is
		indicated by arrow width. The conservation law requires the sum of
		widths to remain constant during backpropagation. Diagram adapted
		from \cite{Heatmapping}.}
	\label{fig:lrpgraph}
\end{figure}

\begin{figure*}
	\centering
	\makebox[\textwidth][c]{\includegraphics[width=1.\textwidth]{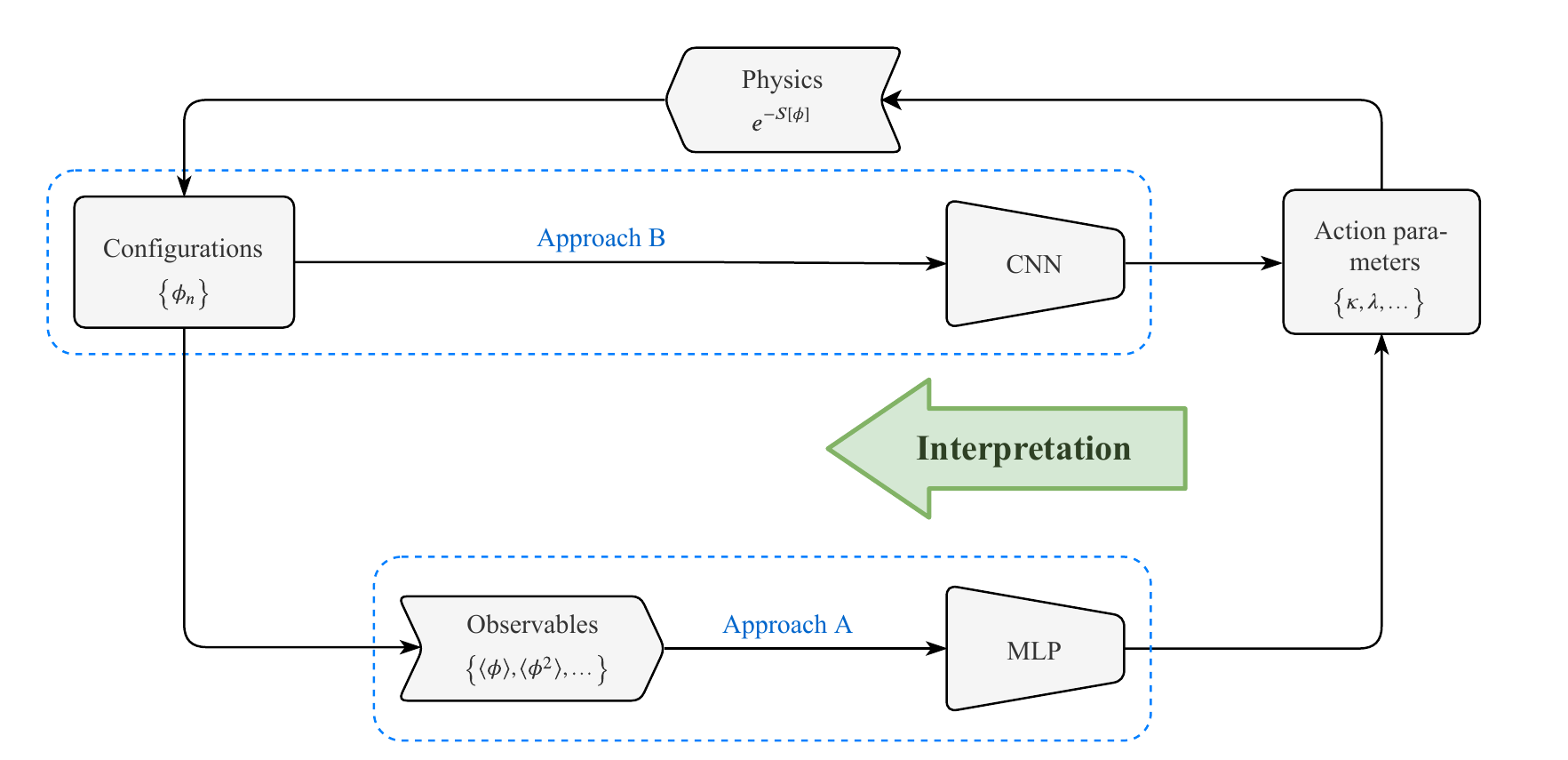}}
	\caption{Sketch of our strategy to learn meaningful structures from
	the simulation data by analysing the networks trained for action
	parameter inference. Field configurations used for training are
	either preprocessed into observables for the MLP (Approach A) or
	directly operated upon with a CNN (Approach B). Obtaining accurate
	predictions for the parameters indicates approximate cycle
	consistency in the above diagram, which supports the notion that the
	networks have successfully identified characteristic features. These
	can then be extracted in a subsequent interpretation step using LRP.}
	\label{fig:RegressionTask}
\end{figure*}

In this work, we focus on LRP, a particular variant of
decomposition-based attribution methods, which has been successfully
applied to other problems in physics and chemistry, e.g.\ in the
context of atomistic systems \cite{Nicoli2018AnalysisOA}.
Nevertheless, we stress that qualitative findings are expected to
agree for all decomposition- and gradient-based methods
\cite{shrikumar2016just}. The general idea of LRP is to start from a
relevance assignment in the output layer and subsequently propagate
this relevance back to the input using certain propagation rules, see
the sketch in \Cref{fig:lrpgraph} and \Cref{app:propagation} for
details. In this way, the method assigns a relevance score to each
neuron, where positive~(negative) entries strongly influence the
classifier towards~(against) a particular classification decision.

\section{Results}
\label{sec:results}

In this section, numerical results are presented which corroborate our
rationale. We train a multilayer perceptron (MLP) and a convolutional
neural network (CNN) to infer the associated hopping parameter
$\kappa$ from a set of known observables (Approach A), as well as
solely from the raw field configurations (Approach B), akin to
\cite{Shanahan2018}. In the first case, without providing any prior
knowledge of the phase boundaries, LRP manages to reveal the
underlying phase structure and returns a phase-dependent importance
hierarchy of the observables in accordance with physical expert
knowledge. In the second case, by calculating the relevances of the
learnt filters, we can associate each of them with one of the physical
phases and thereby extract the known order parameters. Moreover, it
facilitates the construction of an observable that characterises the
symmetric phase. Both variants of our strategy are sketched in
\Cref{fig:RegressionTask}. Due to the ill-conditioned nature of the
action parameter prediction problem, the optimisation objective is
formulated in terms of maximum likelihood estimation. Assuming a
Gaussian distribution with fixed variance, this objective reduces to
minimising the mean squared error (MSE), which we use as loss function
in the following. In addition, we apply weight regularisation, see
\Cref{app:architectures} for details.

\begin{figure}
	\centering
	\includegraphics[width=.97\linewidth]{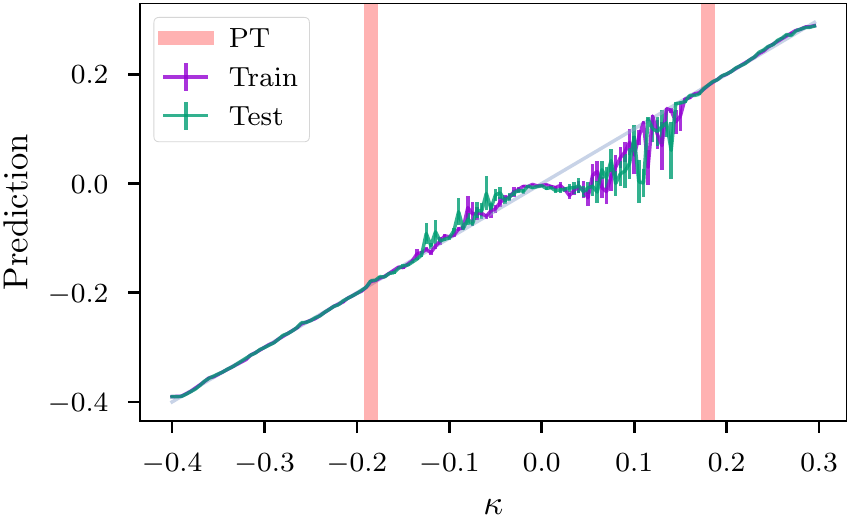}\\
	\vspace{4mm}
	\includegraphics[width=.97\linewidth]{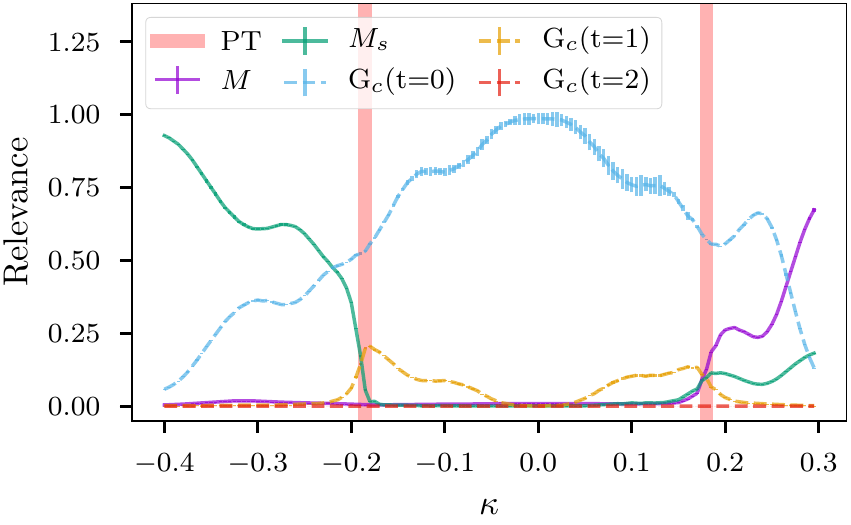}
	\caption{Results for the MLP. Top: predictions, bottom: normalised
	LRP relevances of individual features. Error bars here and throughout
	this work are obtained with the statistical jackknife method.}
	\label{fig:MLP}
\end{figure}
\begin{figure}
	\centering
	\includegraphics[width=.97\linewidth]{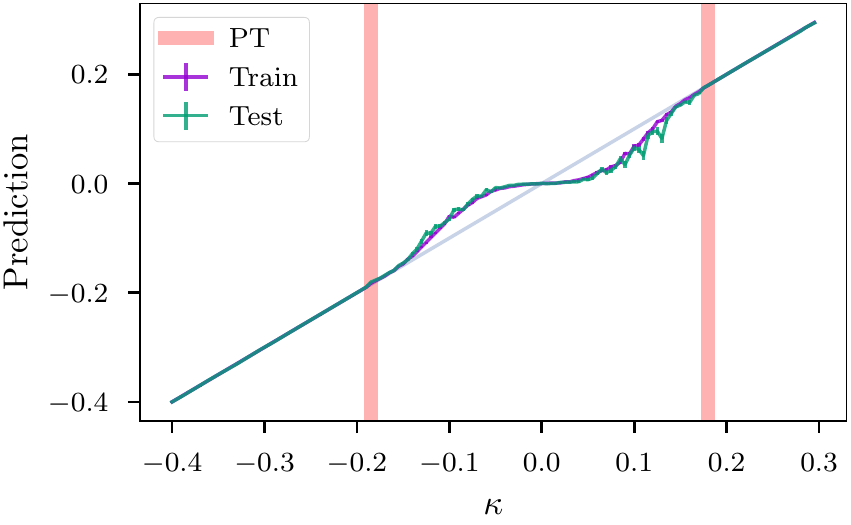}\\
	\vspace{4mm}
	\includegraphics[width=.97\linewidth]{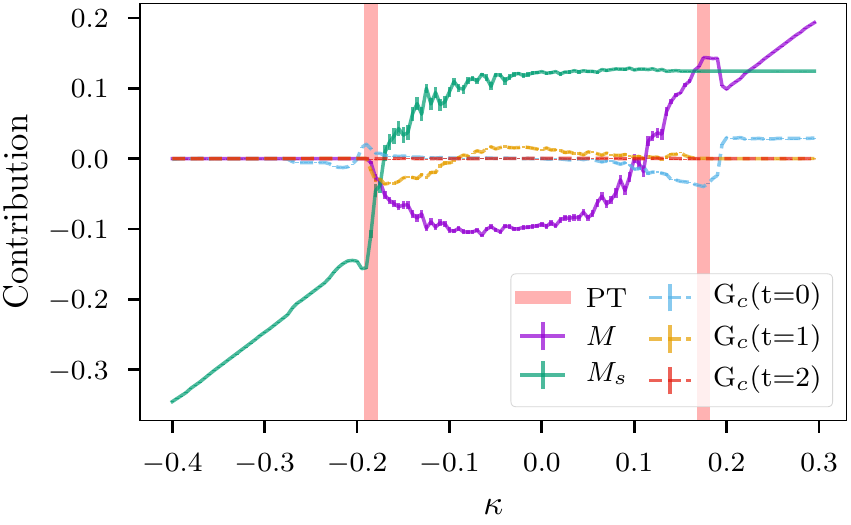}
	\caption{Benchmark results for the random
		forest. Top: predictions, bottom: nominal contributions of
		individual features.}
	\label{fig:RandomForest}
\end{figure}

\subsection{Importance Hierarchies of Known Observables}
\label{sec:exp}

In \Cref{sec:yukawa} we introduced a set of standard observables,
consisting of the normal and staggered magnetisation as well as the
time-sliced two-point correlation function.\footnote{We use a slightly
	modified definition of the time-sliced correlator in order to remove
	lattice artifacts from the data, see \Cref{app:latticedatasets}.} It
seems reasonable to assume that much of the relevant information
characterising the phase structure and dynamics of the theory is
encoded in these quantities. To check this, we create an ordered
dataset of measurements of these quantities at various, evenly spaced
values of $\kappa$ (see \Cref{app:latticedatasets} for details on the
dataset) and use it to perform a regression analysis. We employ a MLP,
also called fully-connected neural network (see
\Cref{app:architectures} for details on the specific architecture).
The method is compared against a random forest regressor as a
baseline, which is a standard method based on the optimisation of
decision trees \cite{Breiman2001} (see \Cref{app:randomforest} for
details). The results for both approaches, shown in
\Cref{fig:MLP,fig:RandomForest}, will be discussed in the following.

We observe qualitatively similar accuracy on the training and test
data in the broken FM and AFM phases. This is expected, since we know
from \Cref{fig:phase} that always one of the two types of
magnetisations is strictly monotonic in the respective phase and can
therefore determine $\kappa$ uniquely. However, both approaches yield
at best mediocre performance in the symmetric PM phase. Here, both
magnetisations tend to zero and therefore do not contain much relevant
information. Moreover, the two-point correlator exhibits approximately
symmetric properties around $\kappa=0$. Therefore, it also does not
provide a unique mapping. This issue is resembled in the prediction
for both methods. The random forest yields a symmetric discrepancy
around $\kappa=0$. In comparison, the MLP shows an improved
performance for $\kappa<0$, albeit at the price of a larger variance
for $\kappa>0$. At this point, we can already see that the chosen set
of observables suffices to characterise the theory only in the broken
phases, whereas in the symmetric phase, additional information appears
to be necessary.

Before we embark on the search for the missing piece, let us first
examine the results further to verify that the learnt decision rules
conform to the physical interpretation given above. We begin with the
relevances as determined by LRP, shown in \Cref{fig:MLP} (bottom), and
later compare to the random forest benchmark below. As expected, $M$
and $M_s$ are relevant in the FM and AFM phases, respectively. There,
considerable relevance is also assigned to the observable $G_c(t=0)$.
However, the contribution appears to diminish when going deeper into
the broken phases. Its comparably large relevance in the symmetric PM
phase shows that it contains most of the information used for the
noisy prediction. As described above, the mediocre performance in this
phase indicates that although the network seems to find weak signals,
the chosen set of observables cannot be optimal.

The interpretation sketched above is further supported by the results
obtained through random forest regression~\cite{Breiman2001}.
Analogously to the previously introduced relevance for LRP, we can
determine nominal contributions of input features to the prediction
and hence a measure of local feature importance (see
\Cref{app:randomforest} for details), which is shown in
\Cref{fig:RandomForest} (bottom). In the broken FM and AFM phases, the
respective contributions of $M$ and $M_s$ demonstrate a linear
dependence on $\kappa$. Again, this clearly indicates that these
quantities characterise the associated phases. For the symmetric PM
phase, the situation appears more challenging, since no such clear
dependence is observed for any of the observables. The non-zero
contributions of features in the PM phase imply that they add some
valuable information to the decision here. However, this has to be
weighted against the observation that the accuracy in this region is
poor. This further confirms our previous conclusion that relevant
information to characterise this phase is largely lost in the
preprocessing step, assuming that it was initially present in the raw
field configurations. It is worthwhile stressing that this analysis
represents an independent confirmation of the results obtained above.
Both algorithms (MLP vs.\ random forest) rely on fundamentally
different principles. We use a model-intrinsic interpretability
measure for the random forest, whereas for the MLP we rely on LRP,
i.e.\ a post-hoc attribution method.

\subsection{Extracting Observables from Convolutional Filters}
\label{sec:extract}

In the previous section, we used a dataset of known observables to
reconstruct $\kappa$. Calculating such quantities corresponds to heavy
preprocessing of the high-dimensional field configuration data. The
resulting low-dimensional features are far less noisy, implying
distillation of relevant information. This is a common procedure in
the field of data science, and may become unavoidable for large
lattices and/or theories with more degrees of freedom. E.g.\ in
state-of-the-art simulations of lattice QCD, the required memory to
store a single field configuration can easily reach
$\mathcal{O}(10^9)$ floating point numbers. Nevertheless, using
preprocessed data in the form of standard observables introduces
strong biases towards known structures. If our perception of the
problem or generally our physical intuition is flawed, machine learning
cannot help us---the relevant information may very well be lost in the
preprocessing step. In the present case specifically, it appears that
important features in the PM phase are neglected by this procedure,
assuming that structures characterising this phase do in fact exist.
Therefore, it is instructive to search for signals of such structures
by training neural networks directly on field configurations.

\begin{figure}
	\centering
	\includegraphics[width=.97\linewidth]{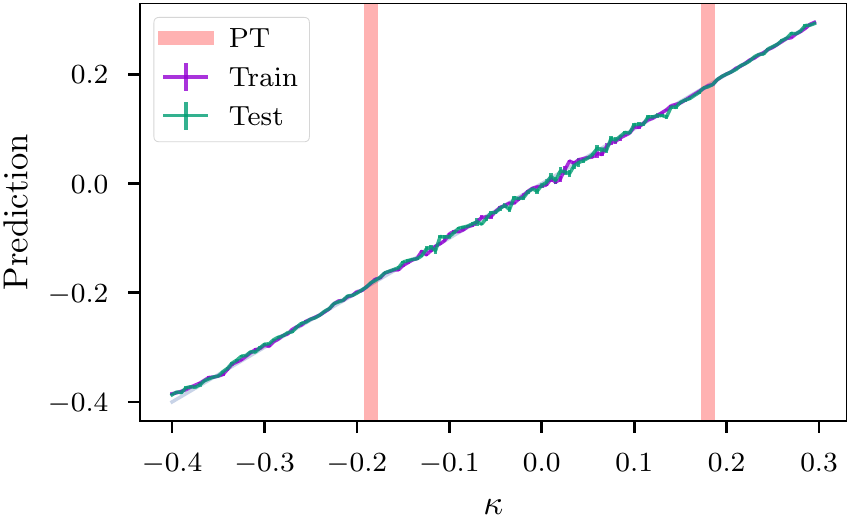}
	\vspace{4mm}
	\includegraphics[width=.97\linewidth]{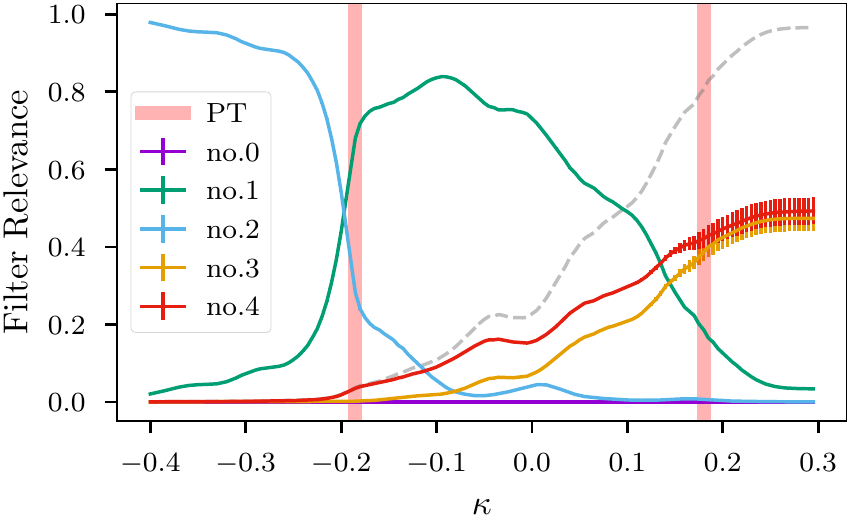}
	\caption{Results for the CNN. Top: prediction, bottom: normalised
	relevances of individual filters. The dashed curve corresponds to the
	cumulative relevance of filter 3 and 4.}
	\label{fig:PredictionCNN}
\end{figure}

As a starting point for this search, we first perform a PCA on the
field configuration dataset. As previously mentioned, this has been
done before with promising results \cite{Hu2017, Wang2016,
	Wetzel2017a}, albeit not in exactly the same physical setting. PCA
immediately identifies the normal and staggered magnetisations as
dominant features, essentially reproducing the work of
\cite{Wetzel2017a}. All higher order principal components show a
vanishing explained variance ratio, implying that no other relevant,
purely linear features are present in the data. This observation
indicates that, if a quantity exists which parametrises the symmetric
PM phase, it cannot simply be a linear combination of the field
variables.

Our improved approach is based on a convolutional neural network
(CNN). The training procedure is largely equivalent to that for the
MLP in the previous section, with the observable dataset replaced by
the full field configurations. We train a CNN using five convolutional
filters with a shape of $2 \times 2 \times 2$ and a stride of $1$. In
order to support explainability, we encourage weight sparsity by
adding the $L^1$ norm to the loss---also known as LASSO
regularisation---as suggested in \cite{Casert2019} (see
\Cref{app:architectures} for details). Due to the nature of the
convolution operation, learnt filters have a direct interpretation
i.t.o.\ first-order linear approximations of relevant observables.
Hence, we expect the CNN to reproduce the PCA results at the very
least, and aim for the identification of other, non-linear quantities,
which the network can encode in subsequent layers. It is important to
understand this difference between the approaches, even though both
extract only linear signals in a first approximation.

The model predictions are shown in \Cref{fig:PredictionCNN} (top). We
can immediately observe a superior performance in the PM phase
compared to our previous results. The CNN succeeds to consistently
infer $\kappa$ from the field configuration data with high accuracy.
This indicates that it indeed manages to construct internal
representations suitable not only to discern the different phases,
which would be sufficient for classification purposes, but also for an
ordering of data points within each phase.

\begin{figure}
	\centering
	\begin{minipage}{0.24\columnwidth}
		\includegraphics[width=.97\linewidth]{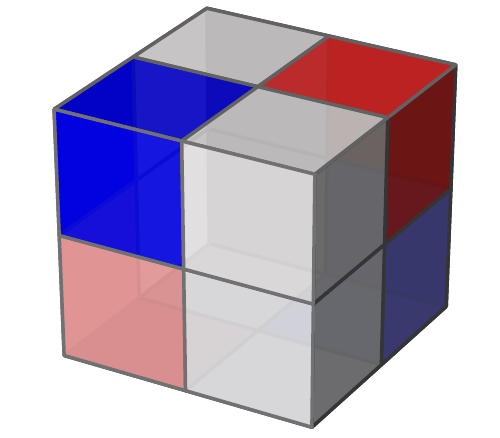}
	\end{minipage}
	\begin{minipage}{0.24\columnwidth}
		\includegraphics[width=.97\linewidth]{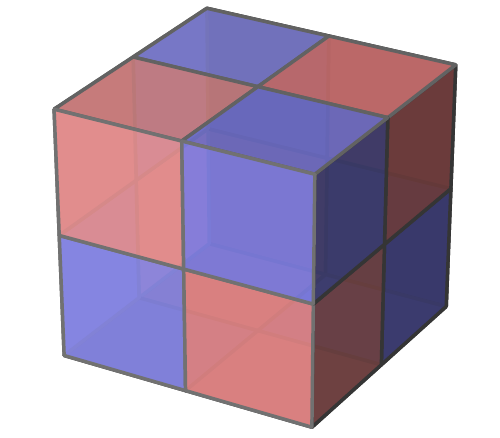}
	\end{minipage}
	\begin{minipage}{0.24\columnwidth}
		\includegraphics[width=.97\linewidth]{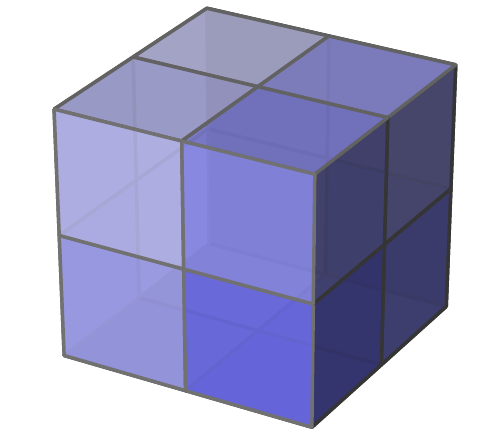}
	\end{minipage}
	\begin{minipage}{0.24\columnwidth}
		\includegraphics[width=.97\linewidth]{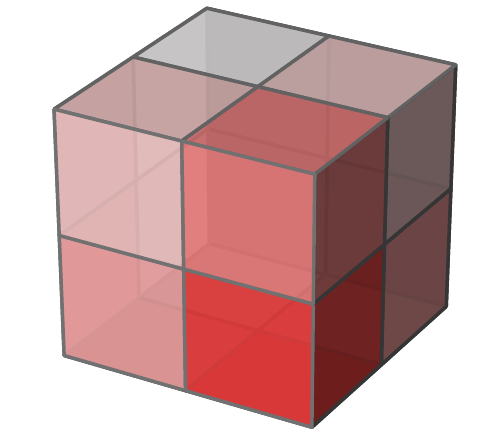}
	\end{minipage}
	\caption{learnt weights of convolutional filters. Left to right:
	(no.1, PM); (no.2, AFM); (no.3, FM); (no.4, FM).
	The colour map is
	symmetric around zero. Red (blue) corresponds to positive (negative)
	weights.}
	\label{fig:FiltersCNN}
\end{figure}

\begin{figure}		
	\begin{minipage}{0.24\columnwidth}
		\includegraphics[width=.97\linewidth]{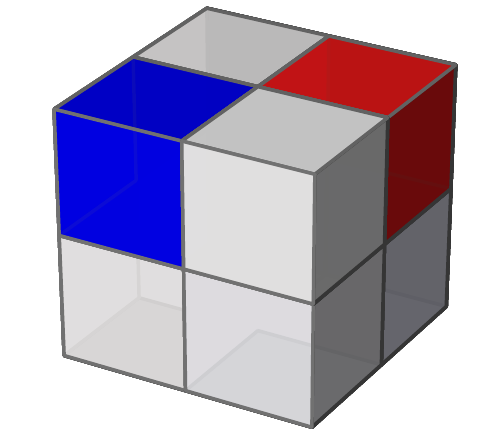}
	\end{minipage}
	\begin{minipage}{0.24\columnwidth}
		\includegraphics[width=.97\linewidth]{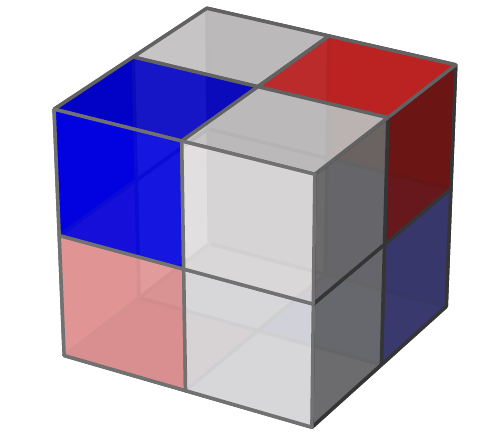}
	\end{minipage}
	\begin{minipage}{0.24\columnwidth}
		\includegraphics[width=.97\linewidth]{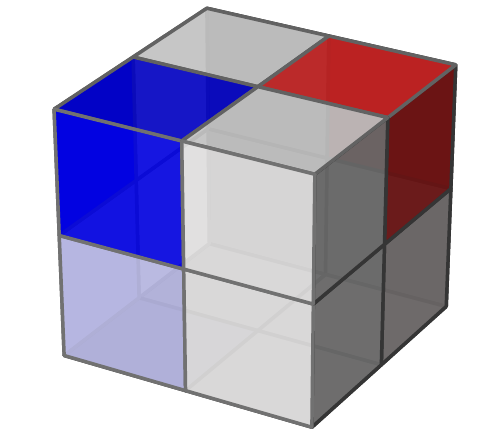}
	\end{minipage}
	\caption{Comparison of PM filters for three independent training runs
	of the CNN.}
	\label{fig:FiltersPM10}
\end{figure}

In order to interpret the predictions and extract knowledge about the
learnt representations, we have to customise LRP to our needs. In
image recognition, as previously mentioned, one mostly aims at
highlighting important regions in the input domain, leading to
superimposed heatmaps. This is based on the inherent heterogeneity
common to image data, where relevant features are usually localised.
For field configurations on the lattice, due to the translational
symmetry of the action and the resulting homogeneity, no particularly
distinguished, localised region should be apparent in any given
sample. However, each convolutional filter encodes an activation map
that is in fact sensitive to a specific feature present in a lattice
configuration. In contrast to the usual ansatz, the spatial
homogeneity promotes global pooling over the relevances associated
with each filter weight. Hence, instead of assigning relevances to
input pixels, we are interested in the cumulative filter relevance
which indicates their individual importance for a particular
prediction. Analogously to the rationale of the previous section, we
can use this approach to build importance hierarchies of filters,
thereby facilitating their physical interpretation as signals of
relevant observables.

\Cref{fig:PredictionCNN} (bottom) shows each filter relevance as a
function of $\kappa$. We can recognise some similarities to the
relevances in \Cref{fig:MLP}, highlighting the underlying phase
structure of the Yukawa theory. It appears that the model can
parametrise each phase individually using one or a small subset of
filters, while the others show small or insignificant relevances in
the respective region. The learnt weight maps are shown in
\Cref{fig:FiltersCNN}, where we also assign names to the filters
depending on the corresponding associated phase, with the exception of
filter no.0 because it exhibits completely vanishing weights and
relevance. It seems to have been dropped entirely by the network,
indicating that four filters are sufficient to characterise all phases
seen in the data. This reduction is an effect of the regularisation,
and constitutes a recurring pattern also when more filters are
initially used, providing a first hint towards the number of
independent quantities utilised by the network.

\begin{figure}
	\centering
	\begin{minipage}{\columnwidth}
		\includegraphics[width=.97\linewidth]{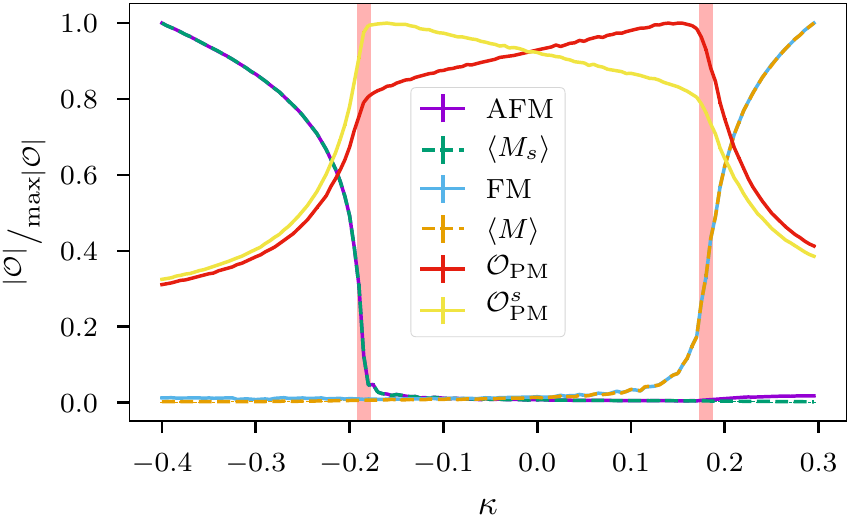}
		\caption{Normalised observables reconstructed from the learnt
		filters. The quantities associated with the FM and AFM phases are
		compared to $M$ and $M_s$. $\mathcal{O}_{\text{PM}}$ and
		$\mathcal{O}_{\text{PM}}^s$ are related by
		\Cref{eq:StaggeredSymmetry} and exhibit an approximate mirror
		symmetry around $\kappa = 0$.} \label{fig:ObservablesReconstructed}
	\end{minipage}
\end{figure}

Let us begin by examining the results that directly correspond to
known quantities. We observe that the FM1 and FM2 filters have entries
of roughly uniform magnitude with a globally flipped sign.
Accordingly, we can identify them as signals of the negative and
positive branches of the magnetisation $M$, respectively. This is
corroborated by their dominating relevances in the FM phase. The AFM
filter exhibits alternating entries of uniform magnitude and therefore
corresponds to the staggered magnetisation $M_s$, which accordingly
dominates the AFM phase. Hence, both order parameters can be
explicitly reconstructed from the CNN. The appearance of two filters
for the magnetisation is easily understood by inspection of the
network architecture in \Cref{tab:NetworkCNN}, the crucial point being
the application of a ReLU activation after the convolution operation.
Consider the action of a positively-valued filter to negatively
magnetised field configurations, or vice versa. The resulting negative
activation map is subsequently defaulted to zero by the ReLU. Hence,
in order to take both branches of $M$ into account, two equivalent
filters with opposing signs are required. The comparably large error
bars in this region stem from the presence of positively and
negatively magnetised samples in the dataset, which lead to a higher
per-filter variance. Therefore, we additionally plot the cumulative
relevance of both filters.

We now discuss the main object of interest, namely the PM filter. It
supplies the dominant signal for the characterisation of this phase.
A linear application of this filter to the configurations, as done for
the FM and AFM filters, does not produce a monotonic quantity, which
would be required for a unique ordering. This further supports the
aforementioned evidence gathered by PCA for the absence of an
additional, purely linear observable. Hence, the simple reconstruction
scheme outlined in the previous paragraphs cannot be applied in this
case. Instead, we undertake a heuristic attempt to reconstruct the
relevant quantity. To this end, we note that the ReLU activation
applied to the convolutional layer's output can effectively correspond
to the absolute value function, albeit with less statistics, if the
entries of the activation map are distributed accordingly. Inspired by
this observation, we define the following observable,

\begin{align}\nonumber
\hspace{-.2cm}\mathcal{O}_\text{PM} &= \frac{1}{|\Lambda|}\sum_{n\in\Lambda}\Big|\,\Bigl[
\phi(n) +\phi(n+\hat{\mu}_1)\Bigr] \\[1ex]
& -\Bigl[\phi(n+\hat{\mu}_2+\hat{\mu}_3) + \phi(n + \hat{\mu}_1 + \hat{\mu}_2 + \hat{\mu}_3)\Bigr]\,\Big|\,.
\label{eq:ObservableReconstructionPM}
\end{align}
As with $M$ and $M_s$, we obtain the corresponding staggered form
$\mathcal{O}_\text{PM}^s$ by applying the transformation given in
\Cref{eq:StaggeredSymmetry}. The resulting pair of quantities is
visualised by the following idealised filters.

\begin{figure}[h!]
	\begin{minipage}{0.24\columnwidth}{	\includegraphics[width=0.97\linewidth]{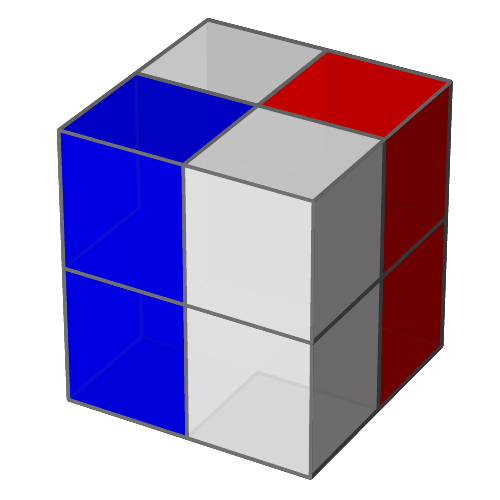}}
	\end{minipage}
	\begin{minipage}{0.24\columnwidth}{	\includegraphics[width=0.97\linewidth]{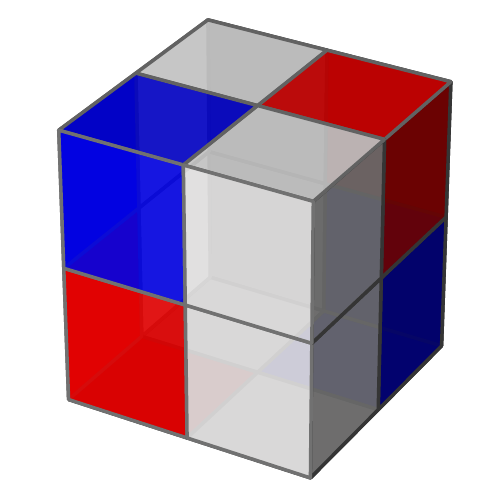}}
	\end{minipage}
	\caption{Convolutional filters corresponding to the observable
		$\mathcal{O}_\text{PM}$ defined in
		\Cref{eq:ObservableReconstructionPM} (left) and its corresponding
		staggered counterpart $\mathcal{O}_\text{PM}^s$ (right).}
\end{figure}

The observable $\mathcal{O}_\text{PM}$ defined in
\Cref{eq:ObservableReconstructionPM} is the sum over all lattice sites
of the lattice derivative in the diagonal $\hat \mu_2+\hat\mu_3$
direction of blocks in the $\hat \mu_1$ direction. This already
explains the modulus, as otherwise $\mathcal{O}_\text{PM}$ would be
the sum over all sites of a total derivative, which vanishes
identically. We also remark that $\mathcal{O}_\text{PM}$ can be made
isotropic by summing over all directions.

We now discuss the properties of the theory that are measured by
$\mathcal{O}_\text{PM}$: In the continuum limit,
$\mathcal{O}_\text{PM}$ naively tends towards the volume integral over
$|\nabla\phi|$. Due to the modulus of the derivative, $\langle
\mathcal{O}_\text{PM}\rangle $ carries the same information as the
expectation value of the kinetic term.

\begin{figure}
	\includegraphics[width=0.4\columnwidth]{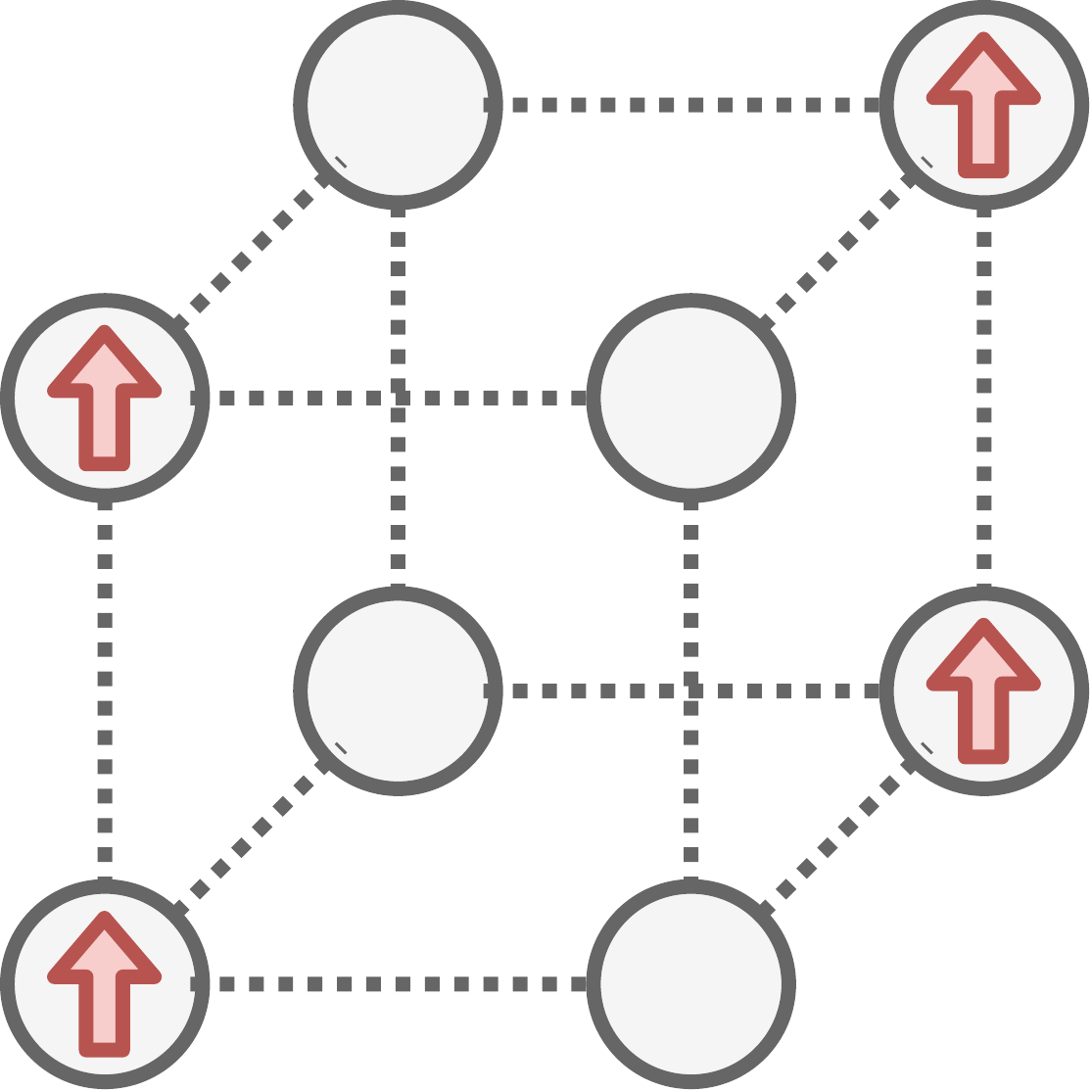}
	\hspace{2ex}
	\includegraphics[width=0.4\columnwidth]{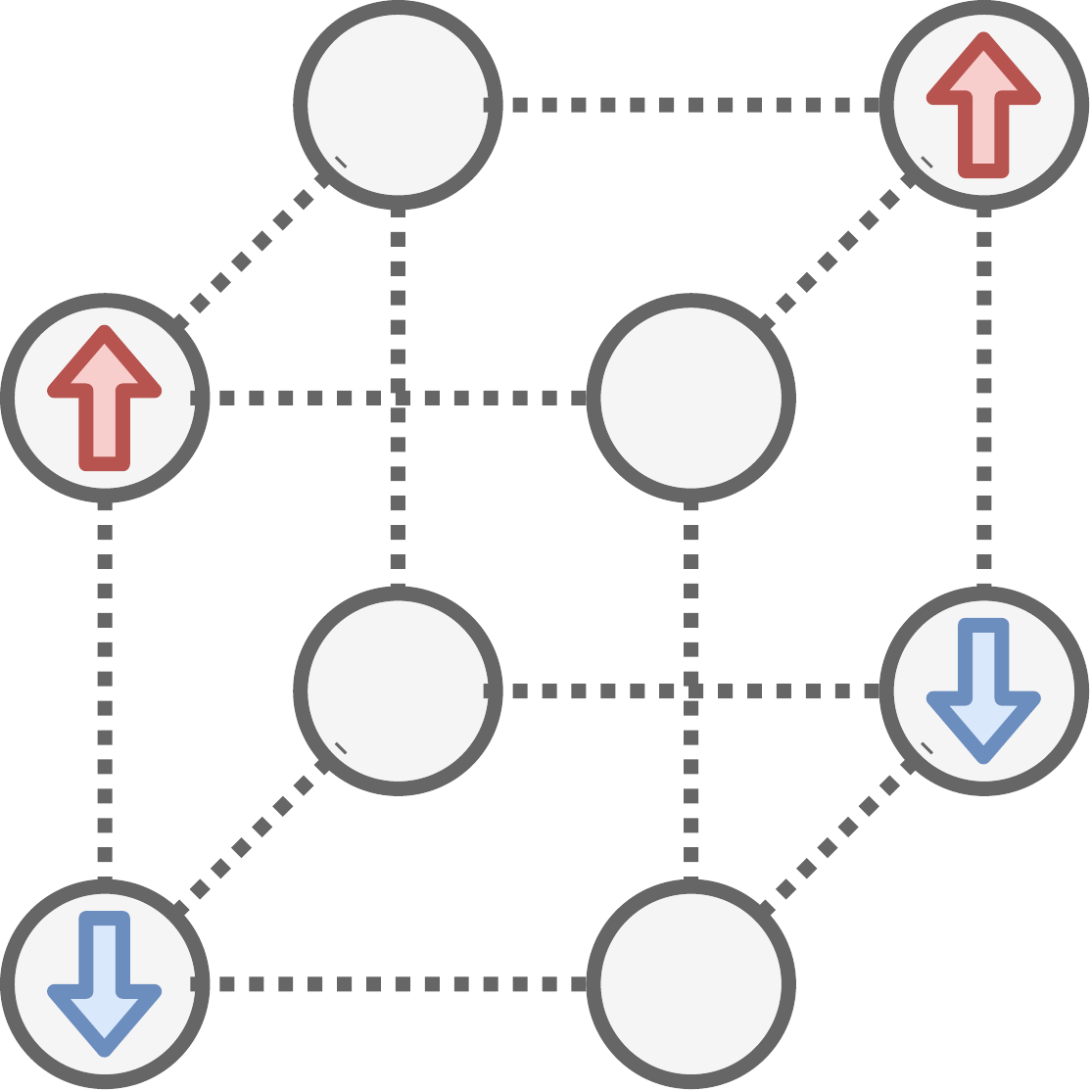}
	\caption{Qualitative visualisation of local structures in field
	configurations operated upon by the PM filter. Sign is encoded by
	arrow orientation/colour. Diagonal neighbours tend to share the same
	sign everywhere in the phase diagram. On the contrary, nearest
	neighbours show a preference towards either same (left) or opposite
	(right) orientations. $\mathcal{O}_\text{PM}$ is particularly
	sensitive towards the local presence/absence of such sign flips in
	the PM phase, without the need for a globally non-zero expectation
	value of the magnetisations.}
	\label{fig:spins}
\end{figure}

The blocking in the $\hat\mu_1$-direction leads to a sensitivity of
$\mathcal{O}_\text{PM}$ to sign flips of  nearest-neighbours. While no
continuum observable is sensitive to these sign flips, the continuum
limit of $\langle \cal{O}_\text{PM}\rangle$ maintains this
information. Accordingly, $\langle \mathcal{O}_\text{PM}\rangle $
exhibits a distinct behavior in the presence of localised,
(anti-)magnetised regions, even if the expectation values vanish
globally. Possible local field alignments resulting in different
values of $\mathcal{O}_\text{PM}$, but not of the standard derivative,
are visualised in \Cref{fig:spins}.

The construction and discussed sensitivities of $\langle
\mathcal{O}_\text{PM}\rangle $ demonstrates again the usefulness of
LRP: we can identify the learnt representation as a feature of the
dataset arising from the lattice discretisation.
$\avg{\mathcal{O}_\text{PM}}$ and $\avg{\mathcal{O}_\text{PM}^s}$ as
functions of $\kappa$ are shown in \Cref{fig:ObservablesReconstructed}
together with the other reconstructed observables and their respective
analytical counterparts. A monotonic, roughly linear dependence is
observed in the PM phase, indicating that the quantity indeed provides
a unique mapping which aids the $\kappa$ inference. In fact, if
$\mathcal{O}_\text{PM}$ is included in the set of predefined
observables for the inference approach detailed in the previous
section, the prediction accuracy of the MLP accordingly becomes
comparable to the CNN in this phase.

In conclusion, we find that the CNN characterises the PM phase by
additionally measuring kinetic contributions in the described manner,
rather than only expectation values of the condensate like in the
broken phases. Still, $M$ and $M_s$ are being utilised as well,
judging from the comparably large relevances of the FM filters in this
region. Due to the opacity of the fully-connected layers following the
convolution, some ambiguity remains regarding the precise decision
rules that the network implements based on these quantities. This
residual lack of clarity can likely be resolved by manually enforcing
locality in the internal operations, e.g.\ by introducing artificial
bottlenecks into the network. Of course, the form of
$\mathcal{O}_\text{PM}$ is also not exactly equivalent to the
operations of the CNN, even though they share many important features.
In particular, there is a mismatch between the averaging procedure and
the MaxPool layer. Effects associated with the choice of different
activation functions and pooling layers, which may be tailored more
specifically towards certain types of observables, should be
investigated in the future. However, our analysis shows that the
overlap with the learned internal representation is significant.

\section{Conclusions and Outlook}
\label{sec:conclusions}

We have investigated the application of interpretability methods to
deep neural network classifiers as a general-purpose framework for the
identification of physical features from lattice data. The approach
facilitates an interpretation of a network's predictions, permitting a
quantitative understanding of the internal representations that the
network learns in order to solve a pretext task---in this case,
inference of action parameters. This culminates in the extraction of
relevant observables from the data, leading to insights about the
phase structure.

First, both types of magnetisations and the time-sliced, connected
two-point correlator were used as training data for a MLP (see
\Cref{fig:MLP}). Inference of the hopping parameter was shown to work
in each of the two broken phases, respectively. However, in the
symmetric phase, the network was observed to suffer from bad accuracy.
This indicates that the amount of relevant information present in the
dataset is insufficient for the network to fully capture the dynamics
of the theory. Using layer-wise relevance propagation, we determined a
$\kappa$-dependent importance hierarchy of the observables. Using this
approach we were able to confirm our physics expectations about order
parameters being relevant within their associated phases. Moreover,
while the two-point correlation function is sensitive to the PM phase,
this signal is insufficient for  attaining high accuracy for the MLP.
Our numerical results and interpretation thereof were further verified
by a random forest regression benchmark performed on the same dataset,
which demonstrated qualitatively comparable accuracy (see
\Cref{fig:RandomForest}).

Next, we trained a CNN directly on the field configurations. In
contrast to aforementioned results, the CNN was shown to yield
superior accuracy for the same inference task (see
\Cref{fig:PredictionCNN}). Therefore, the set of observables chosen
previously must have neglected important information, which the
network managed to distill from the raw data. Employing LRP, a
cumulative relevance was assigned to the individual convolutional
filters, revealing a distinctive pattern that explains the decision
process. In particular, we observed that the network specifically
assigned filters to the each of the phases of the theory, with small
to vanishing relevances in the remaining phases. This also indicates
where phase transitions are located. We confirmed that the learned
filters correspond to representations of the known order parameters by
examining the weight maps (see \Cref{fig:FiltersCNN,fig:FiltersPM10}),
essentially reproducing previous results.

Guided by the filter analysis, we constructed an observable that
characterises the symmetric phase. In a heuristic attempt to find the
exact form of this quantity, we defined $\mathcal{O}_{\text{PM}}$ in
\Cref{eq:ObservableReconstructionPM} and showed that it exhibits
several interesting properties (see
\Cref{fig:ObservablesReconstructed}). We interpreted this quantity as
a particular measure of local fluctuations that is also sensitive to
nearest-neighbour sign flips. This further validates our physical
intuition, since in the PM phase, we expect that relevant information
for its characterisation is encoded by kinetic contributions. As
discussed in detail below \Cref{eq:ObservableReconstructionPM}, the
naive continuum limit of $\mathcal{O}_{\text{PM}}$ is simply the
volume integral of $|\nabla \phi|$, Hence, it has lost the information
about nearest-neighbour sign flips, while the continuum limit of its
expectation value, $\langle \mathcal{O}_{\text{PM}}\rangle $, keeps
its sensitivity towards this property. Accordingly, the construction
of this observable guided by the filter analysis is non-trivial
evidence for the potential power of the present approach: the results
demonstrate that we can identify relevant structures which may
otherwise stay hidden. At this point, LRP has indeed facilitated a
deeper understanding of the CNN, by explaining the origin of its
comparably high accuracy w.r.t.\ the MLP. With these results, we have
conclusively established the value of interpretability methods in deep
learning analyses of lattice data.

In the present work, the emphasis was put on the methodological
aspects of the analysis in order to form a comprehensive basis for
future efforts. Many interesting aspects, such as an investigation of
the fermionic sector, were barely discussed. Instead, we have focused
on the inference of the hopping parameter. Including other action
parameters into the labels, such as the Yukawa coupling or chemical
potential, is a promising endeavour for the future, as it will likely
lead to an improvement in comparison to the current results. This is
necessary in order to pave the way towards an application to more
interesting scenarios, such as QCD at finite density or competing
order regimes in the Hubbard model. Moreover, the introduced ML
pipeline has the potential to provide insight also in various other
areas of computational physics.

\section*{Acknowledgements}
We thank M.~Scherzer, I.-O.~Stamatescu, S.~J. Wetzel and
F.~P.G.~Ziegler for discussions. This work is supported by the
Deutsche Forschungsgemeinschaft (DFG, German Research Foundation)
under Germany's Excellence Strategy EXC 2181/1 - 390900948 (the
Heidelberg STRUCTURES Excellence Cluster) and under the Collaborative
Research Centre SFB 1225 (ISOQUANT), EMMI, the BMBF grants 05P18VHFCA,
01IS14013A (Berlin Big Data Center), and 01IS18037I (Berlin Center for
Machine Learning).

\appendix
\renewcommand{\arraystretch}{1.5}
\setlength{\tabcolsep}{4pt}

\section{Theory and Simulation Details}
\label{app:simulation}

\subsection{Dimensionless Form\\ of the Klein-Gordon Action}

The lattice action for real, scalar $\phi^4$-theory in $d$ dimensions
is defined as

\begin{align}\nonumber
S_\text{KG}[\phi_0] &= \sum_{n\in\Lambda}a^d\Biggr[
\frac{1}{2}\sum_{\mu=1}^{d}\frac{(\phi_{0}(n+a\hat{\mu}) -
	\phi_{0}(n))^2}{a^2}\\
&\hspace{2cm}+ \frac{m_0^2}{2}\phi_0^2 + \frac{g_0}{4!}\phi_0^4\Biggr] \,,
\end{align}
where $a$ is the lattice spacing, $\phi_0, m_0, g_0$ correspond to the
bare field, mass and coupling constant, and $\hat{\mu}$ is the unit
vector in $\mu$-direction. The action can be cast into a dimensionless
form through the following transformation:

\begin{align}\nonumber
a^{\frac{d-2}{2}}\phi_0\quad &= \quad (2\kappa)^{^1/_2}\phi \\ 
(am_0)^2 \quad &= \quad \frac{1 - 2\lambda}{\kappa} - 2d \\
\nonumber
a^{-d+4}\lambda_0 
\quad &= \quad \frac{6\lambda}{\kappa^2}\,.
\end{align}
Here, $\kappa$ is commonly called the hopping parameter and $\lambda$
now takes the role of the coupling constant. Applying this
transformation results in

\begin{align}\nonumber
S_{\text{KG}}[\phi] &=
\sum_{n\in\Lambda}\Biggr[-2\kappa\sum_{\mu=1}^{d}\phi(n)\phi(n+\hat{\mu}) \\
&\hspace{2cm}+ (1- 2\lambda)\phi(n)^2 + \lambda\phi(n)^4\Biggr]\,.
\end{align}

\subsection{Simulating Fermions}

Calculating the determinant of the dicretised Dirac operator
(\Cref{eq:LatticeDiracMatrix}) exactly and repeatedly, which is
in principle necessary for importance sampling, is computationally
intractable even for moderate lattice sizes. The usual approach is to
approximate its value stochastically, e.g.\ by introducing auxiliary
bosonic field variables (commonly called pseudo-fermions), which
guarantees an asymptotically exact distribution. Simulations based on
the numerical solution of differential equations, such as the Hybrid
Monte Carlo (HMC) algorithm or Langevin dynamics, can exploit the
comparably low cost of computing only the matrix inverse with the
conjugate gradient method. In this work, we exclusively employ the HMC
algorithm to generate data.

\section{Lattice Datasets}
\label{app:latticedatasets}

\begin{table}[h]
	\centering
	\begin{tabular}{l |c |c |c |c |l}
		\hline
		$\,N\,$	&	$\,\lambda\,$	&	$\,M\,$	&	$\,g\,$	& $\,\Delta\kappa\,$ &	\#samples per $\kappa$  \\ \hline
		16	&		1.1		&	20	&  0.25 & 0.005 		&	\parbox{4cm}{\vspace{0.15cm} 200 - Training set \\ 100 - Test set \vspace{0.15cm}}	\\
		\hline
	\end{tabular}
	\caption{Action/simulation parameters used for training and test
	dataset.}
	\label{tab:DataSetParameter}
\end{table}

All field configurations composing the datasets used in this work are
generated with the parameters listed in \Cref{tab:DataSetParameter}. A
single, labeled sample is given by the mapping

\begin{equation}
(\phi, \kappa): \qquad\left\{\phi_n\right\} = \left\{\phi_n\mid n\in\Lambda \right\} \longrightarrow \kappa\,.
\end{equation}
In order to explicitly enforce $\mathbb{Z}_2$ symmetry onto the neural
networks, we use the same configurations twice in the dataset, just
with a globally flipped sign. This raw data is directly used to train
the CNN. For the MLP, the samples are preprocessed by computing the
chosen set of observables for each configuration,
 
\begin{equation}
(\mathcal{O}, \kappa): \qquad \{|M|, |M_s|, G_c(t)\} \longrightarrow \kappa\,.
\end{equation}
In this case, we can simply take the modulus of the magnetisations
without losing information, since only two branches with exactly
opposite signs are present in the phase diagram. Due to the finite
expectation value of the staggered magnetisation, the AFM phase
contains unphysical negative correlations. In order to remove these
lattice artifacts, we adapt the usual time-sliced two-point correlator
to

\begin{align}
G_c(t) = \bigg|\avg{\phi(t)\phi(0)} - M^2 - (-1)^{t}M_s^2\bigg|\,.
\label{eq:CorrelationFullCorrection}
\end{align}

Generally, LRP is designed for classification problems. Therefore, we
discretise $\kappa$ to facilitate the formulation of the inference
objective as a classification task. All values of $\kappa$ are
transformed into individual bins and the networks are tasked to
predict the correct bin. In order to retain a notion of locality, the
true bins are additionally smeared out with a Gaussian distribution,
resulting in the target labels

\begin{equation}
\kappa \longrightarrow y_b =  e^{-\frac{\left(\kappa_b-\kappa_\text{True}\right)^2}{2\sigma^2}}\,.
\end{equation}
Here, $b$ denotes the bin number, and the variance was set to $\sigma
= 3 \Delta\kappa$. In combination with a MSE loss, we obtain
qualitatively similar prediction results compared to a standard
regression approach.

\section{Propagation Rules}
\label{app:propagation}

This section contains a summary of the mathematical background of LRP,
in particular regarding the propagation rules. Generally, the
relevance $R_j$ depends on the activation of the previous layer $x_i$.
Given some input to the network, its predicted class $f$ is identified
by the output neuron with the largest response. This neuron's
activation $R^{out}_{f}$, along with $R^{out}_i=0$ for all other
classes $i \neq f$, defines the relevance vector. This output layer
relevance can then be backpropagated through the whole network, which
results in the aforementioned heatmap on the input. Importantly, the
propagation rules are designed such that the total relevance is
conserved,

\begin{align}
\sum_i R^{n}_i = \sum_i R^{out}_i \equiv R_{f}^{out}\,,
\label{eq:lrpConservationLaw}
\end{align}
where the index $n$ can indicate any layer. This conservation law
ensures that explanations from all layers are closely related and
prohibits additional sources of relevance during the backpropagation. A Taylor expansion of this conservation law yields

\begin{align}
\sum_jR_j(x_i) = \underbrace{\sum_jR_j(\widetilde{x}_i)}_{=0} + \sum_{i} \underbrace{\left.\sum_{j} \frac{\partial R_{j}}{\partial x_{i}}\right|_{\left\{\widetilde{x}_{i}\right\}} \left(x_{i}-\widetilde{x}_{i}\right)}_{R_{i}} \, .
\label{eq:lrpTaylorExpansion}
\end{align}
Here, we choose $\widetilde{x}_i$ to be a so-called root point, which
corresponds to an activation with vanishing consecutive layer
relevance $R_j(\widetilde{x}_i)=0$. By definition, it is localised on
the layer's decision boundary, which constitutes a hypersurface in the
activation space. Hence, the root point is not uniquely defined and we
need to impose an additional criterion. However, given such a point,
we can identify the first order term as the relevance propagation rule
$R_j\mapsto R_i$. The remaining root point dependence gives rise to a
variety of possible propagation rules. For instance, the $w^2$ rule
minimises the Euclidean distance between neuron activation $x_i$ and
the decision boundary in order to single out a root point.
Visualisations of root points, as well as essential derivations and
analytical expressions for propagation rules, can be found in
\cite{MonDSP18}.

\section{Random Forest Details}
\label{app:randomforest}

Random forests \cite{Breiman2001} denote a predictive ML approach
based on ensembles of decision trees. They utilise the majority vote
of multiple randomised trees in order to arrive at a prediction. This
greatly improves the generalisation performance compared to using a
single tree. The elementary building block is a node performing binary
decisions based on a single feature criterion. New nodes are connected
sequentially with so-called branches. A single decision tree is grown
iteratively from a root node to multiple leaf nodes. A concrete
prediction corresponds to a unique path from the root to a single
leaf. Each node on the path is associated with a specific feature.
Hence, we can sum up the contributions to the decision separately for
each feature by moving along the path,

\begin{equation}
prediction = bias + \sum_i \left(feature\_contribution\right)_i\,.
\end{equation}
Here, the bias corresponds to the average prediction at the root node.

We employ the \textsc{scikit-learn} implementation \cite{scikit-learn}
in combination with a \textsc{TreeInterpreter} extension
\cite{Saabas2015}. The latter reference also provides an excellent
introduction to the concept of feature contributions.

The random forest is initialised with 10 trees and a maximum tree
depth of 10. This parameter is essential for regularisation, since an
unconstrained depth causes overfitting and thus results in poor
generalisation performance. In order to fix this parameter, we start
at a large value and successively reduce it until the training and
test accuracy reach a similar level. This way we can retain as much
expressive power as possible in the random forest while simultaneously
eliminating systematic errors resulting from overfitting. However, we
emphasise that the specific choice of this parameter not relevant to
our argument.

\section{Network Architectures and Implementation Details}
\label{app:architectures}

\renewcommand{\arraystretch}{1.5}
\setlength{\tabcolsep}{4pt}

We use the \textsc{PyTorch} framework \cite{PyTorch}. The machinery of
LRP is included by defining a custom \texttt{torch.nn.Module} and
equipping all layers with a relevance propagation rule. Furthermore,
all biases are restricted to negative values in order to ensure the
existence of a root point. For training, we employ the Adam optimiser
\cite{Adam} with default hyperparameters and an initial learning rate
of $0.001$, using a batch size of 16.

For both networks, the first layer undergoes least absolute shrinkage
and selection operator (LASSO) regularisation during training, which
encourages sparsity and thereby enhances interpretability. This
corresponds to simply adding the $L^1$ norm of the respective weights
$w_{ij}$ to the MSE loss, which accordingly takes the form

\begin{equation}
L = \frac{1}{d}
\sum_{f=1}^{d}\left(y_{f}-\hat{y}_{f}\right)^{2} +
\lambda_\text{Lasso} \sum_{ij} |w_{ij}|\,.
\end{equation}
Here, $y_f, \hat{y}_f$ denote the prediction and ground truth labels,
and $i,j$ the input and output nodes of the first layer. The quantity
$\lambda_\text{Lasso}$ parametrises the strength of the
regularisation.

The network architectures used in this work are given in the following
tables.

\begin{table}[h!]
	\begin{tabularx}{\columnwidth}{p{2cm} | p{3.5cm} | X}
		\hline
		Layer 			&		Specification	&Propagation rule\\ \hline
		\emph{Linear}	&	in=18, out=256		&$w^2$--\,rule\\
		\emph{ReLU}		&						&$R_i=R_j$	\\
		\emph{Linear}	&	in=256, out=128		&$z^+$--\,rule\\
		\emph{ReLU}		&						&$R_i=R_j$\\
		\emph{Linear}	&	in=128, out=140		&$z^+$--\,rule\\
		\emph{LeakyReLU}&	negative slope=0.01	&$R_i=R_j$\\
		\hline
	\end{tabularx}
	\caption{Network architecture of the MLP. The first layer undergoes
		$L^1$ regularisation with $\lambda_\text{Lasso}=5$.}
	\label{tab:NetworkMLP}
\end{table}

\begin{table}[h!]
	\begin{tabularx}{\columnwidth}{p{2cm} | p{3.5cm}| X}
		\hline 
		Layer			&		Specification								&	Propagation rule			\\ \hline	
		\emph{Conv3d}	&	$\#_\text{filter} = 5\quad\qquad\qquad$ kernel=B, strides=A 	&$w^2$--\,rule\\
		\emph{ReLU}		&													&$R_i=R_j$					\\
		\emph{MaxPool3d}&	kernel=B, strides=B 							&$z^+$--\,rule\\
		\emph{Linear}	&	in=1715, out=256								&$z^+$--\,rule\\
		\emph{ReLU}		&													&$R_i=R_j$\\
		\emph{Linear}	&	in=256, out=140									&$z^+$--\,rule\\
		\emph{ReLU}		&													&$R_i=R_j$\\				
		\hline
	\end{tabularx}
	\caption{Network architecture of the CNN, with $A=(1\times1\times1)$,
		$B=(2\times2\times2)$. The first layer undergoes $L^1$ regularisation
		with $\lambda_\text{Lasso}=10$.  }
	\label{tab:NetworkCNN}
\end{table}

\bibliography{literature}{}	
\bibliographystyle{utphys}

\end{document}